\documentclass[twocolumn,prb]{revtex4}
\usepackage{hyperref}
\usepackage{amsfonts,amsmath,amssymb,graphicx,latexsym}

\begin{document}
\title{Induction and
Mutually Obstructing Equilibria}
\author{Dorothea Hahn}
\author{Mario Liu}\email{mliu@uni-tuebingen.de}
\affiliation{Theoretische Physik, Universit\"{a}t
T\"{u}bingen, 72076 T\"{u}bingen, Germany}
\date{\today}

\begin{abstract}
A unified, consistent and simple view of the Faraday
law of induction is presented, which consists of two
points: discriminating the lab- from the rest-frame
electric field and understanding it is the
impossibility for both fields to vanish
simultaneously, which generates and maintains the
circular current. A number of illustrative examples
are considered, including a mechano-electric pendulum
to exhibit periodic and reversible conversion between
electrical and mechanical energy.
\end{abstract}
\maketitle

\section{Introduction}
\noindent The Faraday law of induction equates the change
of the magnetic flux $\phi$ to the sum of potential drops
along a wire loop,
\begin{equation}\label{1}\textstyle
\frac{{\rm d}}{{\rm d}t}\phi\equiv\frac{{\rm d}}{{\rm
d}t}\int \vec B\cdot{\rm d}\vec A =-\oint\vec E\cdot{\rm
d}\vec \ell=-\sum U_i.
\end{equation}
It contains two effects: The first concerns a changing
field at constant area, $\frac{{\rm d}}{{\rm
d}t}\phi=\int(\frac{\partial}{\partial t}{\vec
B})\cdot{\rm d}\vec A$, and is obtained by integrating
the Maxwell equation, $\frac{\partial}{\partial t}{\vec
B}=-\vec\nabla\times\vec E$.

\begin{figure}[h]
\centering\scalebox{0.25}{\includegraphics{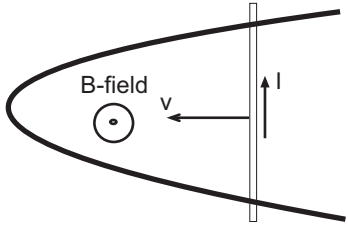}}\\
\caption{A piece of wire moves with the velocity $\vec
v$, changing the area $A$ of the conducting loop in
the presence of a $B$-field. } \label{F1}
\end{figure}

The second effect is given by changing the area of a
conducting loop at a static (and frequently uniform)
field, $\frac{{\rm d}}{{\rm d}t}\phi=\oint\vec
B\cdot(\vec v\times{\rm d}\vec \ell)=\vec
B\cdot\frac{{\rm d}}{{\rm d}t}{\vec A}$. This is a little
harder to grasp: Since $\frac{\partial}{\partial t}{\vec
B}=0$, an integration of the electric field around the
loop is zero, $\int\vec\nabla\times\vec E\cdot{\rm d}\vec
A=\oint\vec E\cdot{\rm d}\vec \ell=0$, and it appears
surprising at first that a current should nevertheless
flow. The prevalent explanation is~\cite{HR}: The
electrons in the moving section of the loop are subject
to the Lorentz force, finite even if the electric field
vanishes, $\vec F=e(\vec E+\vec v\times\vec B)=e\vec
v\times\vec B$. It is their response to, and the
resultant motion along, $\vec F$ that gives rise to the
current $I$, see Fig~\ref{F1}.
More quantitatively, observing that $\vec v\times\vec
B$ is the force per unit charge, same as the electric
field $\vec E$, one concludes that both should also
otherwise be similar. So integrating $\vec v\times\vec
B$ along the loop yields an ``induced" electric
potential $U^{\rm ind}=\oint\vec v\times\vec
B\cdot\,{\rm d}\vec \ell$. And it is only natural to
employ $U^{\rm ind}$ in the Ohm law,
\begin{equation}\label{2}
RI= U^{\rm ind}=\oint\vec v\times\vec B\cdot\,{\rm
d}\vec \ell,
\end{equation}
with $R$ the total electric resistance of the loop, and
$I$ the current. Because this is reminiscent of
batteries, $U^{\rm ind}$ is also referred to as an
electromotive force.

\section{Objections}
When first encountering the Faraday's law, students
are frequently sensitized by their teachers to the
disconcerting fact that this beautifully simple law
lacks a unified understanding, as one needs both the
Lorentz force and one of the Maxwell equations to
derive it. In his famous lectures~\cite{2}, Feynman
succinctly described, and lamented, a general sense of
resignation:
\begin{quote}
We know of no other place in physics where such a
simple and accurate general principle requires for its
real understanding an analysis in terms of {\em two
different phenomena}. Usually such a beautiful
generalization is found to stem from a single deep
underlining principle. Nevertheless, in this case
there does not appear to be any such profound
implication. We have to understand the ``rule" as the
combined effect of two quite separate phenomena.
\end{quote}
Two objections to the above view of induction are also
worthy of some attention. First, there are two different
Lorentz forces: The macroscopic one, $\vec
j\times\langle\vec B\rangle$, given in terms of the
averaged field, expresses the directly veri\-fiable force
on a current carrying wire. The force on an electron,
$e(\vec E+ \vec v_e\times\vec B)$, on the other hand, is
a microscopic formula, given in terms of the electron
velocity $v_{e}$ and the microscopic field $\vec B$.
(Only in this paragraph does $\vec B$ denote the
microscopic field. It is always the macroscopic,
coarse-grained field otherwise.) Electrons in conductors
have a broad velocity distribution, and they are exposed
to strongly varying fields. Therefore, it appears
strikingly bold to assert that the average magnetic force
per unit charge, $\langle\vec v_e\times\vec B\rangle$, is
simply $\vec v\times\langle\vec B\rangle$, the velocity
of the wire times the average $B$-field, as we did above.

It is instructive to reflect upon the Hall Effect in this
context. If evaluated in the same naive fashion,
employing the Lorentz force in a classical free electron
model, the result is notoriously unreliable and rarely
agrees well with experiments. In fact, even the sign may
be wrong --- then it is referred to as the anomalous Hall
effect. So why is the Faraday law universally accurate?

Second, even disregarding these doubts, the above
derivation appears to contain a logical error. Its two
resulting formulas are, for $\vec B$ uniform,
\begin{equation}\textstyle
\vec A\cdot\frac{\partial}{\partial t}{\vec B}=-\sum
U_i, \quad {\vec B}\cdot\frac{{\rm d}}{{\rm d}t}{\vec
A}= -U^{\rm ind}. \label{3}
\end{equation}
We may of course add both equations, obtaining $\vec
A\cdot\frac{\partial}{\partial t}{\vec B}+ {\vec
B}\cdot\frac{{\rm d}}{{\rm d}t}{\vec A}=\dot \phi$ on
the left, with $-U^{\rm ind}$ included in  $-\sum U_i$
on the right, as one usually does. But we must not
forget that each formula remains valid on its own. If
the field is static, $\frac{\partial}{\partial t}{\vec
B}=0$, the sum of potentials vanishes, $\sum U_i=0$,
irrespective whether ${\vec B}\cdot\frac{{\rm d}}{{\rm
d}t}{\vec A}$ vanishes or not. If there is only one
resistive element in the circuit, reducing $\sum U_i$
to a single voltage drop $U_R$, then both this voltage
and the current will always vanish, $I=U_R/R=0$. In
other words, even if ${\vec B}\cdot\frac{{\rm d}}{{\rm
d}t}{\vec A}=-U^{\rm ind}$ is finite, we must not
write $RI=U^{\rm ind}$, as in Eq~(\ref{2}), to account
for Faraday's observation, as this clearly violates
the Maxwell equation.

\section{Two Electric Equilibria}
The consideration below avoids all these difficulties
and inconsistencies. We start by introducing the
electric field $\vec E^0$ of the conductor's local
rest-frame. It is related to the lab-frame field $\vec
E$ as
\begin{equation}\label{4}
\vec E^0=\vec E+\vec v\times\vec B.
\end{equation}
(Only terms to first order in $v/c$ are included in this
paper.) Note that $\vec v$ is the macroscopic velocity of
the medium -- an unambiguous, directly observable
quantity. Rest-frame fields are important, because
conductors strive to reduce them. As long as $\vec E^0$
is finite, there is a current $\vec j=\sigma \vec E^0$,
which redistributes the charge to relax $\vec E^0$ to
zero. Only then is the conductor in equilibrium. One
could not possibly substitute $\vec E$ for $\vec E^0$ in
these statements, because $\vec E$ depends on the
observer's frame that can be changed at will, while the
conductor's equilibrium is an unambiguous fact,
independent of observers.

A metallic object at rest is in equilibrium if $\vec
E=0$; if it moves with the velocity $\vec v$, we have
$\vec E^0=0$ instead, so the lab-frame field is finite,
$\vec E=-\vec v\times\vec B$. In any configuration such
as in Fig.~(\ref{F1}) that offers two inequivalent paths,
``frustration" sets in, as the moving section strives to
establish a finite potential difference, $-\int\vec
v\times\vec B\cdot\,{\rm d}\vec \ell$, by charge
separation, while the stationary part attempts to
eliminate it: The incompatibility of both equilibria,
working hard to obstruct each other, is what gives rise
to a current that flows as long as $\vec v$ is finite.

A limiting case is easy to see: If the resistance of the
sliding bar is much larger than that of the stationary
arc, see Fig.~\ref{F1}, the latter is much better able to
maintain equilibrium, $\vec E\approx0$, so the current
$\vec j=\sigma\vec E_0$ is approximately $\sigma\vec
v\times\vec B$ -- same as obtained above using the
Lorentz force. For a general quantitative account, we
follow Landau and Lifshitz~\cite{LL} to integrate the
Maxwell equation in the form $\frac{\partial}{\partial t}
{\vec B}=-\vec\nabla\times(\vec E^0-\vec v\times\vec B)$,
arriving at
\begin{equation}\label{5}
\textstyle \int{\rm d}\vec
A\cdot\frac{\partial}{\partial t}{ \vec B}+\oint\vec
B\cdot\,(\vec v\times{\rm d}\vec \ell)= -\oint\vec
E^0\cdot\,{\rm d}\vec \ell.
\end{equation}
Identifying the conductor's velocity $\vec v$ with that
the area $\vec A$ changes, the two terms on the left may
be combined as $\frac{{\rm d}}{{\rm d}t}\int \vec
B\cdot{\rm d}\vec A$, and the result is the Faraday's
law, properly given in terms of the rest-frame potential
drops $U^0_i$,
\begin{equation} \label{6}\textstyle
\frac{{\rm d}}{{\rm d}t} \phi\equiv \frac{{\rm d}}{{\rm
d}t}\int \vec B\cdot{\rm d}\vec A= -\oint\vec
E^0\cdot\,{\rm d}\vec \ell=-\sum U^0_i.
\end{equation}

\begin{figure}%
\centering
\scalebox{0.25}{\includegraphics{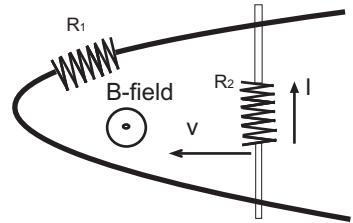}}\\
\caption{Equilibrium requires the rest-frame electric
field of both the stationary and moving section of the
wire to vanish. As this cannot happen simultaneously,
a current is generated ``out of
frustration."}\label{F1a}
\end{figure}

We now revisit Fig.~\ref{F1}, to analyze it as two
inequivalent paths characterized by two resistors, see
Fig.~\ref{F1a}. Clearly, Eq~(\ref{6}) simply states
\begin{equation}\label{6a}\textstyle
\frac{{\rm d}}{{\rm d}t} \phi=-(U_1+U_2)=-(R_1+R_2)I,
\end{equation}
as the result of the general case. Note that the
potential remains constant, $\vec E^0\equiv0$,  between
the resistors, and there is no need for an electromotive
force.

\section{A Mechano-Electric Pendulum}
\begin{figure}
\centering \scalebox{0.25}{\includegraphics{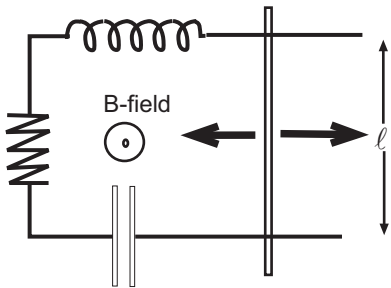}}\\
\caption{A Faraday circuit consisting of  a gliding
wire (mass $M$, length $\ell$), a resistor, a coil and
a capacitor. Wire and coil alone suffice for the
function of a mechano-electric pendulum -- each
standing for one of the two aspects of the Faraday
law. The oscillation of the wire graphically displays
the periodic conversion of electric and mechanical
energy.}\label{F2}
\end{figure}

A worthwhile variation contains all three elements:
resistor, coil and capacitance. With $L$ denoting the
inductivity, $C$ the capacitance,  and $\omega$ the
frequency, they are
\begin{equation}\label{8} U^0_L=-i\omega L\,I,\quad
U^0_C=I/(-i\omega C).
\end{equation}
Inserting them in Eq~(\ref{6}), assuming a uniform and
constant $\vec B$, denoting the relevant length of the
moving wire as $\vec\ell$,  see Fig.~\ref{F2}, we find
\begin{equation}\label{9}\textstyle
[i\omega L-R+(i\omega C)^{-1}]I=\vec B\cdot(\vec
v\times\vec\ell).
\end{equation}
The motion of the wire (of mass $M$) is subject to the
Lorentz force. For uniform $\vec v$, it is given as
\begin{equation}\label{10}\textstyle
M\frac{\partial}{\partial t}{\vec v}=\int\vec j\times\vec
B\,{\rm d}V=\vec\ell\times\vec B I.
\end{equation}
Combining the last two equations, we arrive at a
mechano-electric pendulum,
\begin{equation}\label{11}
\left(\omega^2+i\omega\frac
RL-\omega_0^2\right)I=0,\,\,
\omega_0=\pm\sqrt{\frac1{CL}+\frac{B^2\ell^2}{ML}}.
\end{equation}
Since both the capacitance and the moving wire contribute
to the restoring force, the wire alone would suffice to
form a pendulum with the inductance. Clearly, the numbers
are such that a conveniently observable resonance of
around 1 Hz seems possible -- if the sliding resistance
can be sufficiently reduced~\cite{f}.

\section{Bulk Conductors}
The validity of the Faraday law, Eq~(\ref{6}), is
confined to wire loops, because the conductor's
velocity $\vec v$ was identified with the rate the
area $\vec A$ changes, or $\dot A=v\ell$. Yet the
circumstance of mutually obstructing electric
equilibria, giving rise to currents, occurs under
rather more general conditions -- including especially
bulk metal.
\begin{figure}[b]
\centering \scalebox{0.3}{\includegraphics{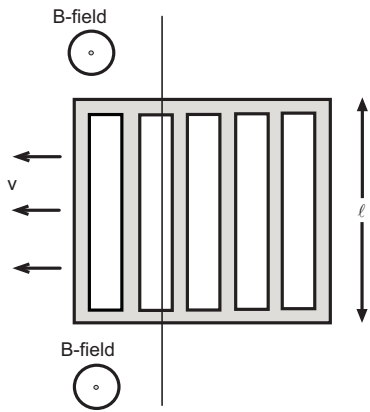}}\\
 \caption{Connected wires moving
 to the left, into the region of a finite magnetic field.}
 \label{F3}
\end{figure}
Consider first the wire grid of Fig.~\ref{F3}, moving
with the velocity $\vec v$ from a region without field
into one with a finite magnetic field. One may of course
apply Eq~(\ref{6}) for all possible loops of this grid,
though it is much simpler to map it onto
Fig.~(\ref{F1a}), considering two inequivalent paths, the
two field-exposed bars striving to establish a potential
difference, and the four field-free ones working to
eliminate it. Each is respectively characterized by the
effective resistance, $R_2=R/2$ and $R_1=R/4$, if we take
all vertical bars to be identical, with resistance $R$,
and neglect the contribution from the horizontal wire.
The result is again Eq~(\ref{6a}), or
\begin{equation}\label{12}
-(R/2+R/4)I=\vec B\cdot(\vec v\times\vec\ell).
\end{equation}
To obtain the breaking force, we insert this
expression for the current $I$ in Eq~(\ref{10}),
\begin{equation}
\textstyle M\frac{\partial}{\partial t}{\vec v}=
-(4\ell^2B^2/3R)\vec v,\qquad\label{13}
\end{equation}
with $3RM/(4\ell^2B^2)$ being the relaxation time --
which of course will change after another vertical
wire moves into the field. If one of the two
horizontal wires is lacking, no current at all will
flow, and no breaking takes place -- because there is
no inequivalent paths. Each portion of the wire net is
happily in equilibrium by itself, without the need or
possibility to obstruct that of the other portion.

Similar circumstances reign when a solid metal plate
enters a magnetic field, because the value of the
effective $R_2$ and $R_1$ decreases with the width of its
region. At the beginning, the field-exposed region is
narrow, and $R_2$ is the dominating one. Toward the end,
when the field-free region is narrow, $R_1$ becomes large
and is the one limiting the current. The largest current
flows, and maximal breaking by the Lorentz force occurs,
when the plate is half in. No current at all flows when
one of the two regions ceases to exist. If a metal comb
enters the field, each tooth is in effect an electrically
independent entity. Due to their narrow width, the
resistance is always large in comparison, so the current,
the Lorentz force and the breaking are always much
smaller.

The next example is the eddy-current break, a piece of
metal moving with $\vec v$, with only part of the
metal exposed to a magnetic field. Equilibrium is
given by $\vec E=0$ outside the field-exposed region,
and by $\vec E=-\vec v\times\vec B$ inside it. Any
deviation from these values churns up a current $\vec
j$ to re-establish them -- obviously quite the same
dilemma as before. The result is again a
frustration-induced current, which dissipates the
kinetic energy of the moving plate effectively.
Constant magnetic field and charge density imply the
field equations:
\begin{equation}\label{14}
\vec\nabla\times\vec E=0, \quad\vec\nabla\cdot\vec
j=\sigma\vec\nabla \cdot(\vec E+\vec v\times\vec B)=0,
\end{equation}
and the boundary conditions: $\triangle E_t=0$,
$\triangle E_n=-\vec v\times\vec B$. These have been
solved~\cite{AJP} assuming constant $\vec B$, $\vec v$,
and a circular field-exposed region. The result is a
dipolar current field, with equal effective resistance,
$R_2=R_1$.

\section{Feynman's Rocking
Contact}
\begin{figure}[h]
\centering \scalebox{0.3}{\includegraphics{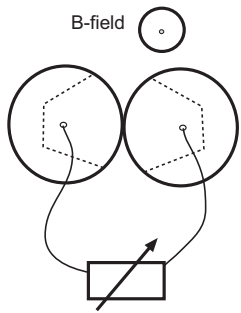}}\\
 \caption{Rocking contacts, as indicated by
 the dotted lines, between
 two plates with curved edges --
 though the physics of the two spherical
 plates should be rather similar.} \label{F4}
\end{figure}
Finally, we discuss Feynman's rocking
contact~\cite{3}, which consists of two metal plates
with slightly curved edges, such that their contact is
only in one point, see the dotted lines of
Fig.~\ref{F4}. It was presented in his lectures as an
example for circumstances in which the Faraday's law
does not hold, with a hint that it could be analyzed
with the Lorentz force. However, in any bulk geometry
such as the one here, there is little reason why the
electron velocity $\vec v_e$ should in any way
resemble $\vec v$ of the metal. So the Lorentz force
cannot be too useful. And it is the rest-frame
electric field that will again do the trick.

To understand the rocking contact, first consider a
metal wheel rotating with the angular velocity
$\omega$, in the presence of a field $\vec
B\,\|\,\vec\omega$. Being in equilibrium, $\vec
E^0=0$, implies $\vec E=-\vec v\times\vec B$ with
$\vec v=\vec\omega\times\vec r_\perp$, or $\vec
E=-\vec r_\perp(\vec\omega\cdot\vec B)$, with the
constant electric density, $\rho=\vec\nabla\cdot\vec
E=-2\vec\omega\cdot\vec B$. Assuming the wheel is
neutral, there is a total surface charge of
$2V\vec\omega\cdot\vec B$ at the rim ($V$ denotes the
volume of the wheel). Reversing the rotation also
reverses the charges.

\begin{figure}
\centering \scalebox{0.3}{\includegraphics{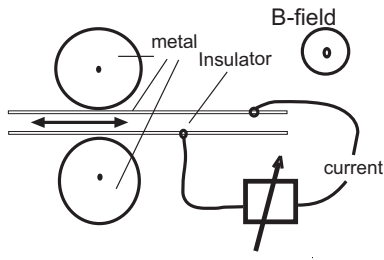}}\\
 \caption{Two metal wheels oscillating in the presence of
 a $B$-field generates an alternating current.}
 \label{F5}
\end{figure}

Next consider Fig.~\ref{F5}, depicting two metal
wheels rotating in place, so the runner in between
moves. The runner has two conducting surfaces
separated by an insulating sheet. Connecting the two
surfaces will give rise to a one-shot current, which
neutralizes the opposite surface charges of the wheels
-- no frustration here. Oscillating the runner
generates an alternating current.

Now reconsider Fig~\ref{F4}, depicting either two
wheels rocking against each other, or if we take away
the metal behind the dotted lines, the rocking contact
of~\cite{3}. The geometry is slightly more complex,
because there is also a contribution from the
translational velocity. But  the basic analysis of the
last example remains valid. Since the rotating parts
are already in contact to each other, little if any
current will travel via the wires.

\section{Summary}
We summarize. Starting from the fact that an electric
equilibrium is given only if the electric field of the
local rest frame vanishes, or $\vec E=-\vec
v\times\vec B$, many phenomena concerning metal parts
moving in the presence of a magnetic field are shown
to become easily understandable, and accessible for
fully or semi-quantitative analysis. The Faraday Law
is seen as a special case of two mutually obstructing
electric equilibria, as a result of which a circular
electric current is maintained.


\begin{thebibliography}{99}
\bibitem{f} It is assumed that the two inductive
elements, say because the fields point in orthorgonal
directions, do not couple.
\end{thebibliography}
\end{document}